\UseRawInputEncoding
\documentclass[
superscriptaddress,
bibnotes,
 amsmath,amssymb,
 aps,
pra,
twocolumn,
longbibliography
]{revtex4-1}
\usepackage{graphicx}
\usepackage{cancel}
\usepackage{dcolumn}
\usepackage{xr}
\usepackage{bm} 
\usepackage{float}
\usepackage{hyperref} 
\hypersetup{
	colorlinks=true,
	linkcolor=blue,
	citecolor=blue,
	filecolor=magenta,
	urlcolor=blue         
}
\usepackage{color,soul}
\usepackage{lineno} 

\renewcommand{\selectlanguage}[1]{}

\begin{document}


\title{Noise and dynamics in acoustoelectric waveguides}

\author{Ryan O. Behunin}
\affiliation{Department of Physics, Northern Arizona University, Flagstaff, Arizona 86011, USA}
\affiliation{Center for Materials Interfaces in Research and Applications, Northern Arizona University, Flagstaff, Arizona 86011, USA}
\email{ryan.behunin@nau.edu}

\author{Andrew Shepherd}
\affiliation{Department of Physics, Northern Arizona University, Flagstaff, Arizona 86011, USA}
\affiliation{Center for Materials Interfaces in Research and Applications, Northern Arizona University, Flagstaff, Arizona 86011, USA}

\author{Ruoyu Yuan}
\affiliation {Department of Electrical Engineering, Yale University, New Haven, Connecticut 06511, USA}

\author{Taylor Ray}
\affiliation{Department of Physics, Northern Arizona University, Flagstaff, Arizona 86011, USA}
\affiliation{Center for Materials Interfaces in Research and Applications, Northern Arizona University, Flagstaff, Arizona 86011, USA}

\author{Matthew J. Storey}
\affiliation{Microsystems Engineering, Science, and Applications, Sandia National Laboratories, Albuquerque, New Mexico, USA}

\author{Peter T. Rakich}
\affiliation {Department of Applied Physics, Yale University, New Haven, Connecticut 06520, USA}

\author{Nils T. Otterstrom}
\affiliation{Microsystems Engineering, Science, and Applications, Sandia National Laboratories, Albuquerque, New Mexico, USA}

\author{Matt Eichenfield}
\affiliation{Microsystems Engineering, Science, and Applications, Sandia National Laboratories, Albuquerque, New Mexico, USA}
\affiliation{James C. Wyant College of Optical Sciences, University of Arizona, Tucson, Arizona 85716, USA}
\affiliation{Electrical, Computer and Energy Engineering, University of Colorado Boulder, Boulder, Colorado 80309, USA}


\date{\today}

\begin{abstract}
We present a quantum field theoretic formulation of acoustoelectric interactions in waveguide-like systems of arbitrary cross-section. Building on an open quantum systems approach, we derive a unified description of plasmon-phonon coupling that incorporates dissipation, noise, and the influence of drift currents. Our analysis captures both bulk and surface plasmon modes, highlighting how drift currents Doppler-shift plasmonic resonances and reshape the phonon noise spectrum. The resulting Heisenberg-Langevin equations yield closed-form expressions for frequency shifts, gain, and noise power spectra, enabling direct evaluation of performance metrics such as the noise factor in acoustoelectric amplifiers and oscillators. In the appropriate limits, this framework reproduces known results while extending them to complex geometries. 
\end{abstract}
\maketitle


\section{Introduction}

In recent years, precision fabrication has enabled phonon-electron couplings that permit powerful new forms of signal processing, mechanical wave amplification, reconfigurable nonlinear optical coupling, and novel oscillator technologies \cite{coldren1971amp,coldren1971wgamp,coldren1973cw,zhou2024electrically,hackett2019amp,hackett2023non,otterstrom2023modulation,wendt2026electrically}. 
At root, these applications leverage the large acoustoelectric couplings that are made possible within complex heterostructures comprised of semiconductors and piezoelectric materials (e.g., see \cite{hackett2024giant}). Despite the remarkable development in new acoustoelectric systems, the theoretical tools that are used to model these systems have not kept pace with these rapid experimental advances. While providing critical insights about device physics and acoustoelectric nonlinearities, established models do not fully capture the impacts of device geometry on acoustoelectric noise and dynamics \cite{kino_normal_1971,kino1973noise,mosekilde1974quantum,hackett2019amp,chatterjee2024ab}. New theoretical tools are needed to maximize the impact of these systems, where a more accurate description of the confinement and dispersion produced in these complex heterostructures may reveal new regimes of operation and functionality.  

Here, we develop an open quantum systems treatment of acoustoelectric interactions that capture the impact of system geometry, dissipation and noise. Our model is based on a Hamiltonian treatment that reproduces the linearized equations describing the conservation of charge, generation of quasistatic electric fields, and charge motion in the presence of a steady drift current. These equations describe two classes of charge oscillation, i.e., bulk and surface plasmon modes. 

For heterostructures where the semiconducting and piezoelectric materials occupy distinct regions of space, we restrict our attention to surface plasmon modes, which generate electric fields that extend beyond the region containing the free charge \cite{barton1979some}. We perform a non-standard form of Sturm-Liouville analysis to identify orthonormal eigenfunctions and eigenfrequencies of these surface modes, and utilize this basis to represent the Hamiltonian in terms of creation and annihilation operators. By coupling each plasmonic degree of freedom to a continuum of bath modes, we derive Heisenberg-Langevin equations that incorporate both dissipation and noise. Finally, piezoelectric couplings between mechanical modes and the electric fields produced by the plasmons, lead to an interaction Hamiltonian with coupling rates determined by acoustoelectric mode overlap.    

To efficiently describe the phonon dynamics in a waveguide, we describe this coupled system in terms of slowly varying envelopes (e.g., see \cite{sipe2016hamiltonian}). This approximation permits a drastic simplification of the dynamics, and is well-adapted to calculate acoustoelectric gain, dispersion, and noise. The calculation of noise is enabled by our open-quantum-systems framework. Because the system operates far from equilibrium, there are no general theorems that can be used to calculate the system's fluctuations \cite{breuer2002theory}. Instead, by specifying the initial state of the bath modes, the Heisenberg-Langevin can be used to compute the noise power spectra for this nonequilibrium system. Our results show that the drift current modifies the phonon noise power spectrum through a Doppler shift of the plasmon oscillation frequencies.    

The paper is organized in the following manner: Sec. \ref{Sec: AEdynamics} shows a Lagrangian formulation of acoustoelectric dynamics, derives the linearized equation of motion, and identifies classes of plasmon modes. The focus of the paper narrows to the Hamiltonian description of the surface plasmons, which have particular relevance to heterostructure-based devices. From a Sturm-Liouville analysis of the surface plasmon equations, orthonormal eigenfunctions and eigenfrequencies are identified. These eigenfunctions form the basis for a normal mode representation of the Hamiltonian. Second quantization provides the quantum dynamics of these surface plasmon modes. In Sec. \ref{Sec: dissipation} the effect of dissipation is modeled. By coupling each surface plasmon mode to a continuum of bath modes, Ohmic damping of the free carrier motion can be described. Using the Lippman-Schwinger orthogonality conditions, this expanded Hamiltonian, including the undamped plasmon interacting with a bath, can be diagonalized. In Sec. \ref{Sec: Envelope}, the acoustoelectric equations of motion are recast in the envelope picture. This formalism is well-adapted to describe slowly-varying changes in phonon amplitude, such as gain or loss.  Sec. \ref{Sec: Heisenberg-Langevin} derives the Heisenberg-Langevin equations for the phonons, capturing the effect of acoustoelectric gain, dispersion and noise. 
In Sec. \ref{Sec: Phonon Noise}, the phonon envelope power spectrum, characterizing the envelope fluctuations, is calculated and used to derive the noise factor for an acoustoelectric amplifier.  
    
\section{Acoustoelectric  dynamics} 
\label{Sec: AEdynamics}

The acoustoelectric effect, involving the interaction of charge oscillations with mechanical motion, can take place within systems that simultaneously support free carrier motion and possess some form of electromechanical coupling (e.g., piezoelectricity) \cite{Parmenter}. In such systems, oscillations of the free charge, i.e., plasmonic modes, can produce electric fields that generate mechanical strain. When set in motion at drift velocity $v_d$, these couplings can mediate complex energy transfer between mechanical vibrations and plasmons. 
\begin{figure*}
    \centering
    \includegraphics[width=0.95\linewidth]{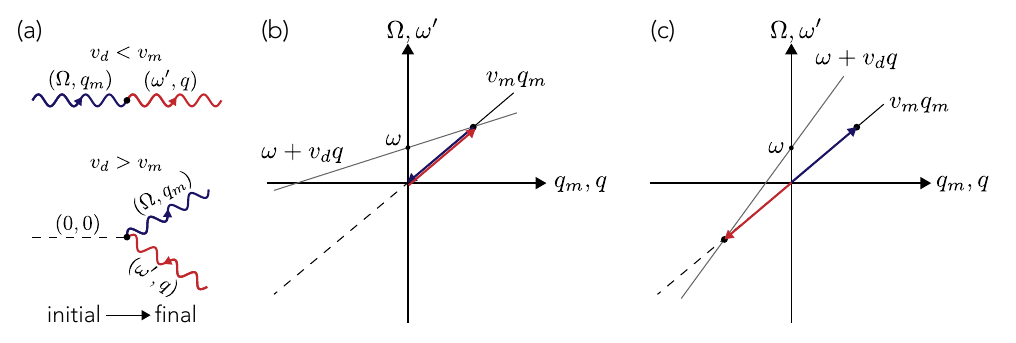}
    \caption{Acoustoelectric scattering processes (a) and phase matching conditions for (b) $v_d<v_m$ and (c) for $v_d>v_m$. Panel (c) shows that when the drift velocity exceeds the speed of sound, a negative lab frame frequency $\omega'$ plasmon can be spontaneously emitted with a phonon. The tails (heads) of the arrows represent the initial (final) state.}
    \label{fig: AE scattering}
\end{figure*}

In the rest frame of the material, a plasmonic mode of frequency $\omega$ and wavevector ${ q}$ oscillates at the Doppler shifted frequency given by $\omega' = \omega + { v}_d  {q}$, where we have restricted our focus to propagation parallel to the drift velocity ${ v}_d$. This plasmon can scatter from a phonon in two ways (Fig. \ref{fig: AE scattering}(a)). The first scattering process can occur in the forward direction, where a phonon of frequency $\Omega$ and wavevector ${q}_m$ (propagating parallel to the drift velocity) is annihilated and a plasmon is created (Fig. \ref{fig: AE scattering}(b)). The phase-matching conditions for this process are given by 
\begin{align}
\label{Eq: PM1a}
    & \Omega  = \underbrace{\omega + {v}_d  {q}}_{\omega'}
   \\
   \label{Eq: PM1b}
   &  { q}_m  =  {q}
\end{align}
and depicted graphically in Fig. \ref{fig: AE scattering}(b).
Using the phonon dispersion relation $\Omega = v_m q_m$, where $v_m$ is the mechanical phase velocity, the phase matching conditions require $\omega = \Omega(1 - v_d/v_m)$. Given that $\omega > 0$, this result shows that this process conserves energy and momentum when $v_m > v_d$. Such forward scattering processes mediate energy transfer from the mechanical field to the free carriers, leading to excess mechanical loss. 

The second process is much like spontaneous parametric down conversion, where phonon-plasmon pairs can be spontaneously emitted from vacuum (lower image in Fig. \ref{fig: AE scattering}(a)), and the drift current plays a role similar to an undepleted pump. This physics can be elucidated by examining the phase-matching conditions for this process 
\begin{align}
\label{Eq: PM2}
    & 0  = \underbrace{\omega + {v}_d  {q}}_{\omega'} + \Omega
   \\
   &  0 =  {q} + { q}_m.
\end{align}
These relations show that the phonon and plasmon are emitted in opposite directions, i.e., the wavevector $q$ is equal and opposite to that of the phonons, i.e., $q = -q_m$. The more interesting consequence comes from Eq. \eqref{Eq: PM2}. Much like the anomalous Doppler effect, this process can only phase match if the plasmon's lab frame frequency is {\it negative}, i.e., $\omega' < 0$, or using Eq. \eqref{Eq: PM2} when $\omega = \Omega(v_d/v_m-1)$ \cite{nezlin1976negative,svidzinsky2019excitation,svidzinsky2021unruh} (Fig. \ref{fig: AE scattering}(c)). Since $\omega$ and $\Omega$ are both positive, we see that this backward scattering process can conserve energy when $v_d > v_m$, or in other words, when the speed of the free carriers exceeds the speed of sound, evoking similarities with Cerenkov radiation. 

To model this physics in waveguide structures, we formulate a quantum field theory of acoustoelectric dynamics within structures of arbitrary, but translationally invariant, cross-sectional geometry. This model is based on the following assumptions; (1) the quasistatic limit is valid, enabling the electric field to be expressed in terms of a scalar potential, (2) the free carriers are well-described using a hydrodynamic description, (3) the electromagnetic dispersion of the material lattice can be neglected, (4) perturbations of the charge density and velocity are well-described to first order, and (5) that the drift current is fixed (i.e., undepleted by acoustoelectric processes). Assumptions (1)-(5) are well satisfied for a broad array of devices.  

We begin from the Lagrangian describing the free carriers and their electrostatic interactions given by 
\begin{widetext}
\begin{align}
\label{L}
    L = \int_{V_{in}} d^3 x  \bigg[ -m (\dot{n} + {\bf v}_{\rm d} \cdot \nabla n) \dot{\psi} &
    - \frac{1}{2} m n_0 (\nabla \dot{\psi})^2 -  e n \varphi + \frac{1}{2} \varepsilon (\nabla \varphi)^2
    \bigg]
    \\
    & + \int_{V_{out}}d^3x \frac{1}{2} \varepsilon (\nabla \varphi)^2
    +  \oint_{\partial V_{in}} da \ \bigg[ 
    -m (\dot{\sigma} + {\bf v}_{\rm d} \cdot \nabla \sigma) \dot{\psi}- e \sigma \varphi \bigg], \nonumber
\end{align}
\end{widetext}
where the displacement of the free carriers from equilibrium is expressed as the gradient of the scalar potential $\psi$ (discussed in more detail below), $n$ is the perturbation of the free-carrier volume density from equilibrium $n_0$, $\sigma$ is the surface charge density,
$m$ and $e$ are the free carrier mass and charge, $\varepsilon$ is the spatially-dependent, but otherwise real and frequency-independent, permittivity of the waveguide structure, and $\varphi$ is the electric potential. Equation \eqref{L} is obtained by adapting a Lagrangian formulation of hydrodynamics \cite{fetter2003theoretical}. For later convenience, the volume integration of the Lagrangian has been explicitly divided into regions that do ($V_{in}$) and do not contain free carriers ($V_{out}$) (Fig. \ref{fig: schematic}). As will be shown below, the term containing ${\bf v}_{\rm d}$ accounts for a constant background drift velocity where the free-carrier drift velocity ${\bf v}_{\rm d}$ is parallel to the direction of translational invariance. 

\begin{figure}
    \centering
    \includegraphics[width=\linewidth]{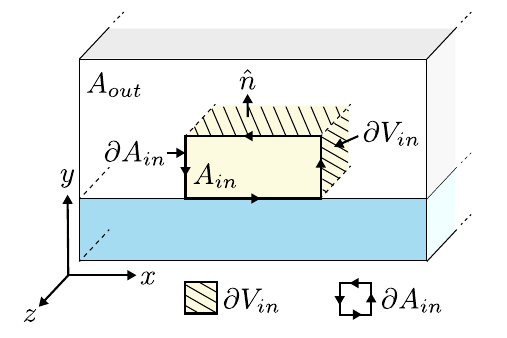}
    \caption{Schematic of acoustoelectric system. The volume $V_{in} (V_{out})$ is the volume formed by $A_{in} (A_{out})$ and the $z$-axis. $A_{in}$ is the cross-section of the region containing the free-carriers.}
    \label{fig: schematic}
\end{figure}

The use of the displacement potential $\psi$ to represent the motion of the free carriers is justified in the quasi-static limit where the electric field is well described by $-\nabla \varphi$. In this regime, the Lorentz force acting on the charged fluid is the gradient of a scalar and, therefore, is curl-free. Defining the displacement of the electrons from equilibrium by ${\bf \xi} = -\nabla \psi$, the linearized (hydrodynamic) equations of motion for the electron fluid about a constant drift velocity ${\bf v}_{\rm d}$ are given by $m \ddot{\bf \xi} + m {\bf v}_{\rm d} \cdot \nabla \dot{\bf \xi} = -e\nabla \varphi$. Taking the curl of both sides of this equation shows $\nabla \times {\bf \xi} = 0$, justifying the representation of the electron displacement by the potential $\psi$. 

To describe the coupling between the potential and the elastic field, we use the interaction Lagrangian given by
\begin{equation}
\label{interaction}
    L_{int} = \int d^3x \ \varphi \partial_i d_{ijk} \partial_j u_k = -H_{int}
\end{equation}
Here, $d_{ijk}$ represents a generic electro-mechanical coupling (e.g, piezoelectric coupling in the stress-charge formulation), $u_j$ is the {\it j}th component of the elastic displacement, and $\partial_k$ represents the $k$th component of the gradient. 

Neglecting the coupling to the elastic field for the moment, the least action principle yields the following equations and boundary conditions
\begin{align}
\label{cont}
& \dot{n} + {\bf v}_{\rm d} \cdot \nabla n  =  n_0 \nabla^2 \dot{\psi}
\\
\label{hydro}
& \ddot{\psi} + {\bf v}_{\rm d} \cdot \nabla \dot{\psi}  =  \frac{e}{m} \varphi
\\
\label{Gauss}
& - \nabla \cdot \epsilon \nabla \varphi  =   e n
\\
\label{sigma}
& \dot{\sigma} + {\bf v}_{\rm d} \cdot \nabla \sigma  =  -n_0 \frac{\partial \dot{\psi}}{\partial x_\perp} 
 \ \ ({\rm for} \ \bf{x} \in \partial V_{in})
\\
\label{BC1}
& -\varepsilon_{out} \frac{\partial \varphi^{out}}{\partial x_\perp} + \varepsilon_{in} \frac{\partial \varphi^{in}}{\partial x_\perp}  =  e\sigma
 \ \ ({\rm for} \ \bf{x} \in \partial V_{in})
 \\
 \label{BC2}
& \varphi^{out}  =  \varphi^{in}
 \ \ ({\rm for} \ \bf{x} \in \partial V_{in})
\end{align}
where $\partial \varphi^{in}/\partial x_\perp (\partial \varphi^{out}/\partial x_\perp)$ and $\varphi^{in}(\varphi^{out})$ is the gradient, normal to the surface of $V_{in}$ (see Fig. \ref{fig: schematic}), and the potential evaluated in the inner (outer surface), respectively. Noting that the perturbation to the free-carrier velocity is given by ${\bf v} = -\nabla \dot{\psi}$, Eqs. \eqref{cont} and \eqref{hydro} reproduce the linearized continuity and hydrodynamic equations in the presence of a uniform drift current, and Eq. \eqref{Gauss} gives Gauss' Law. Equations \eqref{cont}-\eqref{Gauss} can be simplified to obtain effective equations of motion for the charge density, electric potential and free-carrier displacement potential given by 
\begin{equation}
\label{Plasma-Eq}
   ( \nabla \cdot \varepsilon \nabla \partial_\tau^2  
    + \varepsilon_{in} \omega_p^2 \nabla^2 ) 
    \begin{bmatrix}
    \varphi
    \\
    \psi
    \\
    n
    \end{bmatrix}  = 0
\end{equation}
where $\partial_\tau = \partial_t + {\bf v}_{\rm d} \cdot \nabla$, $\partial_\tau^2 = (\partial_t + {\bf v}_{\rm d} \cdot \nabla)^2$, $\varepsilon_{in}$ is the permittivity within the region containing free carriers, and $\omega_p = \sqrt{e^2 n_0/(m \varepsilon_{in})}$ is the permittivity-normalized plasma frequency ($\omega_p = 0$ outside $V_{in}$). Equation \eqref{Plasma-Eq} admits two solution classes: (1) {\it bulk modes} that satisfy $(\partial_\tau^2 + \omega_p^2)\varphi = 0$, and (2) {\it surface modes} that satisfy $-\nabla \cdot \varepsilon \nabla \varphi = 0$ and $\nabla^2 \psi = 0$ everywhere \cite{barton1979some,barton1997van}. Assuming that the drift current flows along the z-direction and that the potential takes the form of a plane wave, i.e., $\varphi \propto \exp\{-i(\omega t -qz)\}$ these modes can be fully characterized using Eqs. \eqref{cont}-\eqref{BC2}. For plane waves, Eqs. \eqref{hydro}, \eqref{sigma} and \eqref{BC1} can be combined to show that the potential must satisfy the Fourier domain boundary condition 
\begin{equation}
\label{SP-BC}
    \varepsilon_{out} \frac{\partial\varphi^{out}}{\partial x_\perp} = \varepsilon_{in}\bigg(1-\frac{\omega_p^2}{\beta^2}\bigg) \frac{\partial\varphi^{in}}{\partial x_\perp}
\end{equation}
where $\beta = \omega - v_{\rm d} q$. For bulk modes, the right-hand side of Eq. \eqref{SP-BC} vanishes, requiring the potential outside $V_{in}$ to vanish \cite{barton1997van}. Consequently, bulk modes do not couple to charges outside of $V_{in}$ for this model and, therefore, for systems where the piezoelectric domain lies outside the semiconducting region, bulk modes do not produce acoustoelectric interactions \cite{barton1979some,barton1997van} (see \footnote{Note however that bulk modes can couple to charges outside of the region containing the free carriers when the effects of diffusion are included.} for an important caveat showing how diffusion modifies the surface and bulk modes). 
Because the piezoelectric materials that mediate electron-phonon coupling are outside the region with free carriers in an array emerging acoustoelectric systems, we focus our attention on the quantum dynamics of surface modes that yield evanescent fields outside $V_{in}$. 

\subsection{Surface mode Lagrangian and Hamiltonian}
To focus our analysis on surface modes, we simplify the system Lagrangian by neglecting the effects of bulk plasmons. Utilizing the divergence theorem, noting that $-\nabla \cdot \varepsilon \nabla \varphi = 0$ and $\nabla^2 \psi = 0$, and $n = 0$ (i.e., $\nabla^2 n = 0$ and $n = 0$ on $\partial V_{in}$ ) we find 

\begin{align}
\label{LS}
 L_S = \oint_{\partial V_{in}} \!\!\!\!da 
     \Bigg[ - & \frac{1}{2}  m  n_0 \dot{\psi} \frac{\partial \dot{\psi}}{\partial x_\perp}
    -m (\dot{\sigma} + {\bf v}_{\rm d}\cdot \nabla \sigma) \dot{\psi} - e \sigma \varphi\nonumber \quad \\
    & +  
    \frac{1}{2} \varphi\bigg(\varepsilon_{in} \frac{\partial \varphi^{in}}{\partial x_\perp}-\varepsilon_{out} \frac{\partial \varphi^{out}}{\partial x_\perp} \bigg) 
     \bigg].
\end{align}
See Appendix \ref{App: variation of surface Lagrangian} for further details on how Euler-Lagrange equations are modified for Lagrangians expressed as surface integrals.

A surface mode Hamiltonian $H_S$ can be derived from Eq. \eqref{LS} by; (1) selecting $\sigma$ to be a generalized coordinate, (2) finding the conjugate momentum to this generalized coordinate $p$, i.e.,  $p \equiv \delta L_S/\delta\dot{\sigma} = - m\dot{\psi}$ (for points restricted to the surface $\partial V_{in}$), and (3) performing a Legendre transform $H_S = \oint_{\partial V_{in}} da \ p \dot{\sigma} - L_S$. This procedure leads to the Hamiltonian for the surface modes given by
\begin{align}
\label{H_S}
 H_S = \oint_{\partial V_{in}} \!\!\!\! da
 \bigg[ &
 \frac{n_0}{2m} p \frac{\partial p}{\partial x_\perp}
 - p {\bf v}_{\rm d}\cdot\nabla \sigma
 + e \sigma \varphi \nonumber \\
 & - \frac{1}{2} \varphi \bigg(
 \varepsilon_{in} \frac{\partial \varphi^{in}}{\partial x_\perp} 
 -\varepsilon_{out} \frac{\partial \varphi^{out}}{\partial x_\perp} \bigg)\bigg].
\end{align}

The dynamics for this coupled system follow from Hamilton's equations as well as the requirement that $-\nabla \cdot \varepsilon\nabla \varphi = 0$ and $\nabla^2 p = 0$, yielding
\begin{align}
\label{EOM-SM1}
 & \dot{\sigma} = \{ \sigma, H_S\} \Rightarrow{} && \partial_\tau \sigma = \frac{n_0}{m} \frac{\partial p}{\partial x_\perp}
 \\
  \label{EOM-SM2}
& \dot{p} = \{ p, H_S\}\Rightarrow{}  && \partial_\tau p = -e \varphi 
 \\
 \label{EOM-SM3}
 & 0 = \frac{\delta H_S}{\delta \varphi} \Rightarrow{} && e\sigma =  \bigg(
 \varepsilon_{in} \frac{\partial \varphi_{in}}{\partial x_\perp} 
 -\varepsilon_{out} \frac{\partial \varphi_{out}}{\partial x_\perp} \bigg),
\end{align}
which are valid for points on the surface of $V_{in}$ and where $\{A,B\}$ is the field-theoretic Poisson bracket associated with generalized coordinate and momenta $\sigma$ and $p$, respectively. This generalized Poisson bracket is defined by 
\begin{eqnarray}
    \{A,B\} \equiv \oint_{\partial V_{in}} d^2 y  \bigg[ \frac{\delta A}{\delta \sigma({\bf y})} \frac{\delta B}{\delta p({\bf y})}- \frac{\delta A}{\delta p({\bf y})}\frac{\delta B}{\delta \sigma({\bf y})} \bigg],
\end{eqnarray}
where Hamilton's equations are reproduced by requiring $\{\sigma({\bf x}),p({\bf x}')\} = \delta^2({\bf x} - {\bf x}')$ where ${\bf x}$ and ${\bf x}'$ lie on the surface $\partial V_{in}$, which follows from the functional derivative definition $\delta \sigma({\bf x})/\delta \sigma({\bf x}') = \delta p({\bf x})/\delta p({\bf x}') = \delta \varphi({\bf x})/\delta \varphi({\bf x}') =\delta^2({\bf x} - {\bf x}')$ (with ${\bf x} \in \partial V_{in}$). 

\subsection{Quantization of surface modes}
Quantization of this system can be achieved by promoting $\sigma$ and $p$ to operators and demanding that the equal time commutation relation (ETCR) $[ \sigma({\bf x}), p({\bf x}')] = i\hbar \delta^2({\bf x}-{\bf x}')$ is satisfied for points on $\partial V_{in}$. Decomposing the surface plasmon system into a collection of normal modes, this postulate shows that the Hamiltonian can be expressed in terms of creation and annihilation operators for surface mode quanta. 

{\it Surface plasmon normal modes:}
Noting that both the electric potential and $p$ satisfy the Laplace equation, we express $\varphi \propto \phi_{\lambda q}({\bf x}_\perp) \exp\{iqz\} A_{\lambda q}$ and $p \propto \phi_{\lambda q}({\bf x}_\perp) \exp\{iqz\} B_{\lambda q}$ where 
$\nabla \cdot \varepsilon \nabla (\phi_{\lambda q}({\bf x}_\perp) \exp\{iqz\}) = 0$ and  $\phi_{\lambda q}({\bf x}_\perp)$ is the eigenfunction of the $\lambda q$-mode. Here, ${\bf x}_\perp$ are the coordinates normal to the $z$-axis, and ${\bf x}_\perp \in \partial A_{in}$ is the set of points bounding the region containing the free carriers (Fig. \ref{fig: schematic}).  

By taking $\varphi \to \phi_{\lambda q}$ and $\beta \to \omega_{\lambda q}$ in Eq. \eqref{SP-BC}, we propose the following modified eigenvalue problem for the surface plasmons 
\begin{eqnarray}
\label{SP_eig_freq}
     -\varepsilon_{out} \frac{\partial\phi^{out}_{\lambda q}}{\partial x_\perp} + \varepsilon_{in}\frac{\partial\phi^{in}_{\lambda q}}{\partial x_\perp} = \frac{\varepsilon_{in} \omega_p^2}{\omega_{\lambda q}^2}\frac{\partial\phi^{in}_{\lambda q}}{\partial x_\perp}
\end{eqnarray}
where $\omega_{\lambda q}$ is the oscillation frequency of the $\lambda q$th mode. Using this equation, as well as the Laplace equation, we can show that these functions form a complete basis that can be used to represent any function on the surface  $\partial V_{in}$ and that the surface plasmon frequencies $\omega_{\lambda q}$ are real. 

Using the plane-wave decomposition permitted by translation invariance in the $z$-direction, the Laplace equation becomes 
\begin{eqnarray}
\label{eig_func}
&& \nabla_\perp \cdot \varepsilon \nabla_\perp \phi_{\lambda q} = \varepsilon q^2 \phi_{\lambda q}.
\end{eqnarray}
Multiplying on the left by $\phi^*_{\lambda' q'}$ and integrating over a planar region $S$ that is perpendicular to the $z$-direction, we find
\begin{eqnarray}
    \int_{S} da \ \phi^*_{\lambda' q'}\nabla_\perp \cdot \varepsilon \nabla_\perp \phi_{\lambda q} = q^2 \int_{S} da \ \varepsilon \ \phi^*_{\lambda' q'} \phi_{\lambda q}
\end{eqnarray}
Using the divergence theorem, the Laplacian operator can be transferred onto $\phi^*_{\lambda' q'}$, providing the relationship given by  
 \begin{align}
 \label{SL}
         \oint_{\partial S} d\ell \ \varepsilon \ \bigg(
        \phi^*_{\lambda' q'}  \frac{\partial \phi_{\lambda q}}{\partial x_\perp}
        - & \frac{\partial \phi^*_{\lambda' q'}}{\partial x_\perp}   \phi_{\lambda q} \bigg) = 
        \\
        &   (q^2-q'^2) \int_{S} da \ \varepsilon \ \phi^*_{\lambda' q'} \phi_{\lambda q}. \nonumber 
 \end{align}
 Equation \eqref{SL} provides two important results when $q=q'$. First, we find 
 \begin{eqnarray}
 \label{SL2}
        \oint_{\partial S} d\ell \ \varepsilon \         \phi^*_{\lambda' q}  \frac{\partial \phi_{\lambda q}}{\partial x_\perp}
         = \oint_{\partial S} d\ell \ \varepsilon \ \frac{\partial \phi^*_{\lambda' q}}{\partial x_\perp}   \phi_{\lambda q}.
 \end{eqnarray}
 Second, taking the sum of Eq. \eqref{SL} for $S$ given by the cross section of $V_{in}$, $A_{in}$, and $S$ given by the entire $xy$-plane, excluding $A_{in}$ we find
 \begin{align}
 \label{SL3}
        \oint_{\partial A_{in}} d\ell \ 
        \Bigg[  
        \phi^*_{\lambda' q}  
        & \bigg(
         -\varepsilon_{out}\frac{\partial \phi^{out}_{\lambda q}}{\partial x_\perp}
         +\varepsilon_{in} \frac{\partial \phi^{in}_{\lambda q}}{\partial x_\perp}\bigg)
        -
        \\
        & \bigg(
         -\varepsilon_{out}\frac{\partial \phi^{out*}_{\lambda' q}}{\partial x_\perp}
         +\varepsilon_{in} \frac{\partial \phi^{in*}_{\lambda' q}}{\partial x_\perp}\bigg) \phi_{\lambda q}\bigg]
        = 0. \nonumber
 \end{align}
 where continuity of $\phi_{\lambda q}$ at $\partial A_{in}$ has been used. Utilizing Eqs. \eqref{SP_eig_freq} and \eqref{SL2}, Eq. \eqref{SL3}, after some simplification, yields the modified orthogonality conditions for the surface plasmon modes given by
\begin{eqnarray}
 \label{orth_rel-1}
       (\omega_{\lambda'q}^{*2}-\omega_{\lambda q}^{2}) \oint_{\partial A_{in}} d\ell \ 
            \phi^*_{\lambda' q}  
          \frac{\partial \phi^{in}_{\lambda q}}{\partial x_\perp}
        = 0.
 \end{eqnarray}
 For $\lambda = \lambda'$, we show in Appendix \ref{App: orthogonality} that the integral $\oint_{\partial A_{in}} d\ell \  \phi^*_{\lambda q}
 \frac{\partial \phi^{in}_{\lambda q}}{\partial x_\perp}$ is real and positive definite,  requiring that $\omega_{\lambda q}$ is real valued. For $\lambda \neq \lambda'$, and assuming that the eigenfrequencies $\omega_{\lambda q}$ are non-degenerate, Eq. \eqref{orth_rel-1} shows that $\phi^*_{\lambda' q}$ are orthogonal to $\partial \phi^{in}_{\lambda q}/\partial x_\perp$ on the boundary of $A_{in}$. Assuming that these eigenfunctions are normalized, we obtain the orthogonality conditions 
\begin{eqnarray}
 \label{orth_rel-2}
      \oint_{\partial A_{in}} d\ell \  
            \phi^*_{\lambda' q}  
          \frac{\partial \phi^{in}_{\lambda q}}{\partial x_\perp}
          = 
          \oint_{\partial A_{in}} d\ell \  
          \frac{\partial \phi^{*in}_{\lambda' q}}{\partial x_\perp} \phi_{\lambda q}  
      & = & \delta_{\lambda \lambda'}.  \ \ 
 \end{eqnarray}


\subsection{Normal mode decomposition of fields and Hamiltonian}
Using the ETCR, the Heisenberg equations of motion can be derived, taking the same form as Eqs. \eqref{EOM-SM1}-\eqref{EOM-SM3}. When decomposed into normal modes of amplitude $a_{\lambda q}$ that satisfy the commutation relations
\begin{eqnarray}
\label{Eq:CR-1}
    && [a_{\lambda q}, a^\dag_{\lambda' q'}] = \delta_{\lambda \lambda'} \delta(q-q'),
    \\
    \label{Eq:CR-2}
    && [a_{\lambda q}, a_{\lambda' q'}] = [a^\dag_{\lambda q}, a^\dag_{\lambda' q'}] = 0,
\end{eqnarray}
we find that the following field expansions satisfy the ETCR and the Heisenberg equations of motion  
\begin{eqnarray}
 && 
 \label{phi-SME}
 \varphi =  \sum_\lambda \int \frac{dq}{\sqrt{2\pi}}   \frac{e \sqrt{n_0} \omega^2_{\lambda q} }{\varepsilon_{in} \omega_p^2}  \phi_{\lambda q} e^{iqz} x_{\lambda q}\quad  
 \\
&& 
\sigma = \sum_\lambda \int \frac{dq}{\sqrt{2\pi}}  \sqrt{n_0}  
 \frac{\partial \phi^{in}_{\lambda q}}{\partial x_\perp} e^{iqz}  x_{\lambda q}  \quad  
 \\
&& p = \sum_\lambda \int \frac{dq}{\sqrt{2\pi}}\frac{1}{\sqrt{n_0}} 
 \phi_{\lambda q} e^{iqz} p_{\lambda q}  \quad
\end{eqnarray}
where 
\begin{eqnarray}
    x_{\lambda q} = && \sqrt{\frac{\hbar}{2 m \omega_{\lambda q}}}(a_{\lambda q} + a^\dag_{\lambda,-q}),
    \\
    p_{\lambda q} = && -i \sqrt{\frac{\hbar m \omega_{\lambda q}}{2 }}(a_{\lambda q} - a^\dag_{\lambda,-q}),
\end{eqnarray}
and, without loss of generality, we have assumed that $\phi_{\lambda q}$ is a real-valued function. Using Eqs. \eqref{Eq:CR-1} and \eqref{Eq:CR-2}, the commutation relations for $x_{\lambda q}$ and $p_{\lambda q}$ give 
\begin{eqnarray}
\label{Eq:CR-xp}
    [x_{\lambda q},p_{\lambda' q'}] = i \hbar \delta_{\lambda \lambda'} \delta(q+q').
\end{eqnarray}

Inserting these mode expansions for $\varphi$, $\sigma$ and $p$ into Eq. \eqref{H_S}, and using the orthonormality relations and the eigenvalue equation, the Hamiltonian for the surface plasmon modes reduces to 
\begin{align}
\label{H_S_NM}
    H_S = & 
   \sum_\lambda\! \int dq  \bigg(
   \frac{p_{\lambda q}^\dag p_{\lambda q}}{2m} \!+\! iv_d q \ p_{\lambda q} x^\dag_{\lambda q} 
  \! + \! \frac{m \omega_{\lambda q}^2}{2} x^\dag_{\lambda q} x_{\lambda q} \bigg)\nonumber
   \\
   = & \sum_\lambda \int dq \ \frac{1}{2} \hbar( \omega_{\lambda q} +v_d q) (a^\dag_{\lambda q} a_{\lambda q}+a_{\lambda q} a^\dag_{\lambda q}) .
\end{align}
Noting that $p_{\lambda q}^\dag = p_{\lambda,- q}$ and $x_{\lambda q}^\dag = x_{\lambda,- q}$, $H_S$ can be shown to be Hermitian.

\subsection{Phonon Hamiltonian}
The phonon Hamiltonian is given by 
\begin{eqnarray}
    H_{ph} = \sum_{\mu} \int dq \ \hbar\Omega_{\mu q}b^\dag_{\mu q} b_{\mu q}
\end{eqnarray}
where $\mu$ denotes the mode family and $q$ denotes the wavevector for propagation along the $z$-direction. Here, $b_{\mu q}$ and $b^\dag_{\mu q}$ are phonon annihilation and creation operators.

This Hamiltonian follows from the quantization of 
\begin{eqnarray}
    H_{ph} = \int d^3x \ \bigg[
    \frac{1}{2\rho}\vec{\Pi}^2+\frac{1}{2}C_{ijkl} \partial_i u_j \partial_k u_l
    \bigg]
\end{eqnarray}
where $u_k$ is the $k$th component of the elastic displacement, $\vec{\Pi}$ is the conjugate momentum to ${\bf u}$, $\rho$ is the (generally spatially-dependent) mass density, and $C_{ijkl}$ denotes the components of the elastic stiffness tensor. Hamilton's equations for $H_{ph}$ reproduce the equations of linear elasticity.

Demanding that $[u_j({\bf x}), \Pi_k({\bf x}')] = i \hbar \delta_{jk} \delta^3({\bf x}-{\bf x}')$ (i.e., the ETCR for the phonons), the fields ${\bf u}$ and $\vec{\Pi}$ can be decomposed into normal modes
\begin{align}
      & {\bf u} = \sum_\mu \int dq \ \sqrt{\frac{\hbar}{4\pi \Omega_{\mu q}}}\bigg[ 
    \vec{\mathcal{U}}_{\mu q} e^{iqz} b_{\mu q} + {\rm H.c.} 
    \bigg]
    \\
   & \vec{\Pi} = -i\sum_\mu \int dq \ \sqrt{\frac{\hbar \Omega_{\mu q}}{4\pi }} \rho\bigg[ 
    \vec{\mathcal{U}}_{\mu q} e^{iqz} b_{\mu q} - {\rm H.c.} 
    \bigg]
\end{align}
where $\vec{\mathcal{U}}_{\mu q}e^{iqz}$ is an eigenfunction of the elastic equation $\partial_i C_{ijkl} \partial_k (\mathcal{U}_{\mu q,l}e^{iqz}) = -\rho \Omega_{\mu q}^2 (\mathcal{U}_{\mu q,j}e^{iqz})$ that satisfies the orthonormality relation $\int d^3 x \rho \ \vec{\mathcal{U}}^*_{\mu' q'}\cdot
\vec{\mathcal{U}}_{\mu q}e^{i(q-q')z}
= \delta_{\mu \mu'} (2\pi)\delta(q-q')$ and $\mu$ is a generic label denoting the different modes of the system. The ETCR for the phonon field require
\begin{eqnarray}
   &&  [b_{\mu q}, b^\dag_{\mu',q'}] = \delta_{\mu \mu'} \delta(q-q')
    \\
  &&   [b_{\mu q}, b_{\mu' q'}] =  [b^\dag_{\mu q},b^\dag_{\mu',q'}] = 0.
\end{eqnarray}

\subsection{Interaction Hamiltonian}
With plasmon and phonon fields expressed in terms of normal modes, the interaction Hamiltonian (Eq. \eqref{interaction}) can be expressed in terms of creation and annihilation operators. Utilizing the representation of the fields in terms of normal modes, $H_{int}$ is given by 
\begin{eqnarray}
    H_{int}  = 
    - \sum_{\lambda,\mu}\int dq \ \hbar \bigg[
    g_{\lambda\mu q}(a_{\lambda,-q} +a^\dag_{\lambda,q}) b_{\mu q} + {\rm H.c.}\bigg]
\end{eqnarray}
where the coupling rate $g_{\lambda\mu q}$ is given by 
\begin{eqnarray}
\label{Eq: AE coupling}
    g_{\lambda\mu q}  &=& \sqrt{\frac{\omega_{\lambda q}^3}{4 \varepsilon_{in} \omega_p^2 \Omega_{\mu q}}} \int da \ \phi_{\lambda q} \tilde{\partial}^*_m d_{ijm} \tilde{\partial}^*_i \mathcal{U}_{\mu q,j}.
\end{eqnarray}
Here, $da = dx dy$ and $\tilde{\partial}_i$ is the $i$th component of the gradient where the $z$-component is given by $\tilde{\partial}_z = i q$.

\section{System dynamics including Ohmic damping}
\label{Sec: dissipation}

Losses are critical for the accurate description of the acoustoelectric effect. To model these effects, we utilize techniques of open quantum systems. By coupling each plasmon mode to a continuous collection of bath modes, the effective system dynamics, where the bath degrees of freedom are integrated out, exhibit loss and noise \cite{breuer2002theory}. With the appropriate choice of system-bath coupling (discussed below), Ohmic damping can be realized. Such an open system treatment is necessitated because of the over-damped nature of plasmons in many acoustoelectric systems. In this limit, we find that simple white noise models do not capture the appropriate thermal properties of the acoustoelectrically coupled phonons.  

To establish these open-system dynamics, we consider each plasmon mode individually. 
Focusing on the $\lambda q$-mode, the Hamiltonian $H_{\lambda q}$ is given by 
\begin{align}
    H_{\lambda q} = & \frac{p_{\lambda q}^\dag p_{\lambda q}}{2m} +  iv_d q \ p_{\lambda q} x^\dag_{\lambda q} 
   + \frac{1}{2} m \omega_{\lambda q}^2 x^\dag_{\lambda q} x_{\lambda q} + H_{\lambda q}^{bath} 
\end{align}
where the term $H_{\lambda q}^{bath}$ given by 
\begin{align}
    H_{\lambda q}^{bath} = 
   &  \int_0^\infty d\nu \bigg[
    \frac{p_{\lambda q \nu}^\dag p_{\lambda q \nu}}{2m} +i v_d q \ p_{\lambda q \nu} x^\dag_{\lambda q \nu}   
    \\
    & + \frac{1}{2} m\nu^2(x_{\lambda q \nu}^\dag - c_\nu x_{\lambda q}^\dag)(x_{\lambda q \nu} - c_\nu x_{\lambda q})
    \bigg]
    , \nonumber
\end{align}
accounts for the dynamics of the `bath', and the bilinear coupling between the bath modes and plasmons (i.e., this coupling does not capture nonlinear effects). To yield Ohmic dissipation, the coupling coefficient is given by 
\begin{eqnarray}
    c_\nu = \sqrt{\frac{2\gamma}{\pi}} \frac{1}{\nu} \equiv \frac{1}{\nu^2} G_\nu.
\end{eqnarray}
Here, $\gamma$ is the free-carrier scattering rate (plasmon decay rate), and the scaled coupling parameter $G_\nu =\nu \sqrt{2\gamma/\pi}$ has been introduced for later convenience. 

A straightforward but lengthy calculation (see Appendix \ref{App: diagonalization}) shows that the following operators diagonalize $H_{\lambda q}$
\begin{align}
\label{Eq:Sys-x}
  &  x_{\lambda q} =  \int_0^\infty \! d\omega  \sqrt{\frac{\hbar}{2m\omega}}[\chi_{\lambda q}(\omega) a_{\lambda q \omega } + \chi_{\lambda q}^*(\omega) a_{\lambda,- q, \omega }^\dag]
    \\
    \label{Eq:Sys-p}
   & p_{\lambda q} =  -i\int_0^\infty \! d\omega  \sqrt{\frac{\hbar m \omega}{2}} [\chi_{\lambda q}(\omega) a_{\lambda q \omega } - \chi_{\lambda q}^*(\omega) a_{\lambda,- q, \omega }^\dag]
    \\
 &   x_{\lambda q \nu} =  \int_0^\infty \! d\omega  \sqrt{\frac{\hbar}{2m\omega}}[\chi_{\lambda q\nu}(\omega) a_{\lambda q \omega } + \chi_{\lambda q\nu}^*(\omega) a_{\lambda,- q, \omega }^\dag]
    \\
    \label{Eq: bath-p}
   & p_{\lambda q \nu} =  -i\int_0^\infty \! d\omega  \sqrt{\frac{\hbar m \omega}{2}} [\chi_{\lambda q\nu}(\omega) a_{\lambda q \omega } - \chi_{\lambda q\nu}^*(\omega) a_{\lambda,- q, \omega }^\dag]
\end{align}
where the $a_{\lambda q \omega }$ and $a^\dag_{\lambda q \omega}$ satisfy the commutation relations
\begin{eqnarray}
    && [a_{\lambda q \omega},a^\dag_{\lambda' q' \omega'}] = \delta_{\lambda \lambda'}\delta(q-q')\delta(\omega-\omega') \\ 
    && [a_{\lambda q \omega},a_{\lambda' q' \omega'}] = 0 \\ 
    && [a^\dag_{\lambda q \omega},a^\dag_{\lambda' q' \omega'}] = 0.
\end{eqnarray}
Here, $\chi_{\lambda q}(\omega)$ and $\chi_{\lambda q\nu}(\omega)$ are susceptibility functions given by 
\begin{eqnarray}
\label{Eq: susc-1}
   && \chi_{\lambda q}(\omega) = \frac{G_\omega}{-\omega^2 - i \gamma \omega + \omega_{\lambda q}^2}
   \\
   \label{Eq: susc-2}
   && \chi_{\lambda q\nu}(\omega) = \delta(\nu-\omega) + \frac{G_\nu}{\nu^2-(\omega + i 0^+)^2} \chi_{\lambda q}(\omega) \quad
\end{eqnarray}
that satisfy the Lippmann-Schwinger orthogonality relations
\begin{align}
\label{Eq: LS-1}
    & \int_0^\infty d\omega \chi_{\lambda q}^*(\omega) \chi_{\lambda q}(\omega) = 1
    \\
    \label{Eq: LS-2}
    & \chi_{\lambda q}^*(\omega) \chi_{\lambda q}(\omega') +  \int _0^\infty  d\nu \ \chi_{\lambda q \nu}^*(\omega) \chi_{\lambda q \nu}(\omega')  =  \delta(\omega-\omega') 
    \\
    \label{Eq: LS-3}
    & \int _0^\infty d\omega \ \chi_{\lambda q \nu}^*(\omega) \chi_{\lambda q \nu'}(\omega) = \delta(\nu-\nu')
\\
\label{Eq: LS-4}
    &
    \int_0^\infty d\omega \ \chi_{\lambda q \nu}^*(\omega) \chi_{\lambda q}(\omega) = 0.
\end{align}
We obtain the open systems Hamiltonian $H$ by inserting Eqs. \eqref{Eq:Sys-x}-\eqref{Eq: bath-p} into $H = \sum_\lambda \int dq \ H_{\lambda q}$ 
\begin{eqnarray}
    H =  \sum_\lambda \int dq \int_0^\infty d\omega \ \hbar(\omega + v_d q) a^\dag_{\lambda q \omega} a_{\lambda q \omega}
\end{eqnarray}
where a divergent zero-point energy has been discarded. Neglecting the zero-point energy amounts to resetting the zero of energy and does not impact the system dynamics. Finally,  with the use of Eqs. \eqref{Eq:Sys-x}-\eqref{Eq: bath-p}, the total Hamiltonian including the interaction with the phonons as well as the phonon evolution is given by
\begin{widetext}
\begin{eqnarray}
\label{Eq: open system H}
    H+H_{ph}+H_{int} = && \sum_\lambda \int dq \int_0^\infty d\omega \ \hbar(\omega + v_d q) a^\dag_{\lambda q \omega} a_{\lambda q \omega}
    + 
    \sum_{\mu} \int dq \ \hbar\Omega_{\mu q}b^\dag_{\mu q} b_{\mu q}  \nonumber
    \\
    && - \sum_{\lambda,\mu}\int dq \int_0^\infty d\omega \ \hbar \sqrt{\frac{\omega_{\lambda q}}{\omega}}\bigg[
    g_{\lambda\mu q}\big(\chi_{\lambda q}(\omega) a_{\lambda,-q,\omega} +\chi^*_{\lambda q}(\omega)a^\dag_{\lambda q\omega}\big) b_{\mu q} + {\rm H.c.}\bigg].
\end{eqnarray}
\end{widetext}

See Appendix \ref{App: diagonalization} for more details on the derivation of $H$. 

\section{Envelope formulation}
\label{Sec: Envelope}
Here, we develop the envelope formulation of acoustoelectric interactions, providing a well-adapted framework to describe spatial growth (or decay) of the phonon fields within waveguide structures. The central assumption for this approximation is that the spatial rate of change in the amplitude is much smaller than the carrier wavevector $q_m$, or more precisely, the spatial dynamics of the fields is well-described by a narrow band of wave-vectors (interestingly, these assumptions can fail in emerging systems \cite{hackett2019high}). Owing to the phase-matching conditions given by Eqs. \eqref{Eq: PM1a} \& \eqref{Eq: PM1b}, this traveling-wave phonon can resonantly couple with right-moving ($+$) or left-moving ($-$) plasmons. 
The envelope operators are defined by 
\begin{eqnarray}
    B_\mu(z) & = & \frac{1}{\sqrt{2\pi}}\int dq \ e^{-i(q_m-q)z} b_{\mu q} 
    \\
    \Phi_{\lambda \omega \pm}(z) & = & \frac{1}{\sqrt{2\pi}}\int dq \ e^{-i(\pm q_m-q)z} a^{(\pm)}_{\lambda q \omega}, 
\end{eqnarray}
where it is assumed that $b_{\mu q}$ is sharply peaked near the dominant phonon wavevector $q_m$, and $a^{(\pm)}_{\lambda q \omega}$ represents the amplitude for a band of plasmons with wavevector peaked near $q \approx \pm q_m$. Note that the hybridized system + bath amplitudes $a_{\lambda q \omega}$ entering the plasmon envelope definition freely evolve {\it losslessly} as they describe a closed system, justifying the envelope approximation above. We note that Hamiltonian Eq. \eqref{Eq: open system H} permits a plasmon-mediated coupling between forward- and backward-propagating phonons. However, this coupling is suppressed in the regime of validity of the rotating wave approximation (e.g., weak-coupling and relatively high mechanical Q) and is neglected from hereon. 

Utilizing the commutation relations for $a_{\lambda q\omega}$ and $b_{\mu q}$, one can show
\begin{align}
& [B_\mu(z),B^\dag_{\mu'}(z')]  = \delta_{\mu \mu'} \delta(z-z') \\
& [B_\mu(z),B_{\mu'}(z')]  =  0
\\
& [\Phi_{\lambda\omega\pm}(z),\Phi^\dag_{\lambda'\omega'\pm}(z')]  = \delta_{\lambda \lambda'} \delta(z-z') \delta(\omega-\omega')
    \\
    & [\Phi_{\lambda\omega\pm}(z),\Phi_{\lambda'\omega'\pm}(z')]  =  [\Phi^\dag_{\lambda\omega\pm}(z),\Phi^\dag_{\lambda'\omega'\mp}(z')]  \!=\!  0.
\end{align}
Here, we have assumed that forward and backward moving plasmons (i.e., $q \approx \pm q_m$) are sharply peaked enough that correlations between $\Phi_{\lambda+}(z)$ and $\Phi_{\lambda-}(z)$ can be neglected. 

In the envelope formulation, the Hamiltonian becomes
\begin{widetext}
\begin{align}
\label{H_env}
    H =  \hbar \int dz \bigg\{ 
    \sum_{\lambda,\pm} \int_0^\infty d\omega \ \Phi^\dag_{\lambda \omega\pm} &\hat{\omega}_{\pm} \Phi_{\lambda\omega\pm}
    + 
    \sum_{\mu}  \ B^\dag_{\mu}  \hat{\Omega}_{\mu} B_{\mu} \nonumber
    \\
    & -  \sum_{\lambda,\mu} \int_0^\infty d\omega \sqrt{\frac{\omega_{\lambda q_m}}{\omega}} \bigg[g_{\lambda \mu q_m} \bigg( \chi_{\lambda q_m}(\omega)
    \Phi_{\lambda \omega-}
    + \chi^*_{\lambda q_m}(\omega)\Phi^\dag_{\lambda \omega+}\bigg) B_\mu + {\rm H.c.}
    \bigg]\bigg\}
\end{align}
\end{widetext}
where the frequency operators are defined by 
\begin{eqnarray}
  &&  \hat{\omega}_{\pm} =\omega \pm v_dq_m - iv_d \frac{\partial}{\partial z}
    \\
 &&   \hat{\Omega}_{\mu} = \sum_{n=0}^\infty \frac{1}{n!}
    \frac{\partial^n \Omega_{\mu q}}{\partial q^n} \bigg|_{q=q_m}\bigg(-i\frac{\partial}{\partial z}\bigg)^n,
\end{eqnarray}
and spatial dispersion of the coupling rate has been neglected, i.e., $g_{\lambda \mu q} \approx g_{\lambda \mu q_m}$. 
For slowly varying phonon envelopes, the frequency operator can be well-approximated by retaining the first few terms
\begin{eqnarray}
 &&   \hat{\Omega}_{\mu} \approx \Omega_{\mu q_m}-i v_{\mu q_m}\frac{\partial}{\partial z}.
\end{eqnarray}
where 
$v_{\mu q_m}$ is the group velocity of the $\mu$th phonon mode. 

\section{Heisenberg-Langevin equations}
\label{Sec: Heisenberg-Langevin}

With the Hamiltonian established, we now derive the equations for the phonon dynamics that capture gain and noise. 
We begin by calculating the Heisenberg equations of motion for the phonon and plasmon envelopes. Then, narrowing our focus to a single phonon mode, we neglect the sums on $\mu$ and suppress the labels $\mu$ and $q_m$ (e.g., $v_{\mu q_m} = v_g$, $\omega_{\lambda q_m}\to \omega_\lambda$, $g_{\lambda \mu q_m}\to g_\lambda$, $\Omega_{\mu q_m} \to \Omega$, $B_\mu \to B$), leading to the simplified Heisenberg equations
\begin{widetext}
\begin{eqnarray}
\label{Eq: HOM-phonon}
    && \dot{B} = 
    -i\Omega B - v_g \partial_z B+ i \sum_\lambda \int_0^\infty d \omega \ \sqrt{\frac{\omega_{\lambda}}{\omega}}g^*_{\lambda} \big(\chi^*_{\lambda}(\omega)\Phi^\dag_{\lambda \omega -}
    +  \chi_{\lambda}(\omega)\Phi_{\lambda \omega +} \big)
    \\
    \label{Eq: plasmon_right}
   && \dot{\Phi}_{\lambda \omega +} =  -i(\omega+v_d q_m) \Phi_{\lambda \omega +}-v_d\partial_z \Phi_{\lambda \omega +}+ i  \sqrt{\frac{\omega_{\lambda}}{\omega}} g_{\lambda} \chi^*_{\lambda}(\omega) B
    \\
    \label{Eq: plasmon_left}
  &&  \dot{\Phi}_{\lambda \omega -} =  -i(\omega-v_d q_m) \Phi_{\lambda \omega -}-v_d\partial_z \Phi_{\lambda \omega -}+ i \sqrt{\frac{\omega_{\lambda}}{\omega}} g^*_{\lambda} \chi^*_{\lambda}(\omega)B^\dag.
\end{eqnarray}
By inserting the formal solution for the plasmon envelopes in the equation for $B$, effective equations of motion for the phonon envelope can be obtained that capture the gain and noise produced by the acoustoelectric effect. 

The solution for Eqs. \eqref{Eq: plasmon_right} and \eqref{Eq: plasmon_left} are given by 
\begin{eqnarray}
&&   
\label{Eq: plasmon_right_sol}
\Phi_{\lambda +}(t,z) = \hat{\Phi}_{\lambda +}(t,z) + i \sqrt{\frac{\omega_{\lambda}}{\omega}} g_{\lambda}  \chi^*_{\lambda}(\omega) \int^{\infty}_0 d\tau \ e^{-i(\omega+v_d q_m)\tau} B(t-\tau,z-v_d\tau)
    \\
&& 
\label{Eq: plasmon_left_sol}
\Phi_{\lambda -}(t,z) = \hat{\Phi}_{\lambda -}(t,z) + i \sqrt{\frac{\omega_{\lambda}}{\omega}} g_{\lambda }^* \chi^*_{\lambda}(\omega) \int^{\infty}_0 d\tau \ e^{-i(\omega-v_d q_m)\tau} B^\dag(t-\tau,z-v_d\tau) 
\end{eqnarray}
where $\hat{\Phi}_{\lambda +}(t,z)(\hat{\Phi}_{\lambda -}(t,z))$ is the homogeneous solution for the plasmon envelopes, capturing the noise imparted to the phonons by the acoustoelectric effect. 

The Heisenberg-Langevin equation for the phonon is derived by inserting the solutions for $\Phi_{\lambda \pm}$ into Eq. \eqref{Eq: HOM-phonon}.   
\begin{eqnarray}
\label{Eq: HL-phonon-2}
 \dot{B}(t,z) = 
    -i\Omega B(t,z) - v_g \partial_z B(t,z) 
    + i 2\sum_\lambda \int_0^\infty d\tau  \frac{\omega_\lambda}{\tilde{\omega}_\lambda}|{g}_\lambda|^2   
    e^{-(iv_dq_m+\gamma/2)\tau} \sin (\tilde{\omega}_\lambda \tau)B(t-\tau,z-v_d \tau) + \xi_{AE}(t,z).
\end{eqnarray}
Here, $\tilde{\omega}_\lambda \equiv \sqrt{\omega_\lambda^2 -\gamma^2/4}$ and the result has been simplified by evaluating an integral over frequency (Appendix \ref{Sec: Simplification HE}). The source term $\xi_{AE}$ is a Langevin force that captures the plasmon fluctuations that are imparted to the phonon
\begin{align}
   \xi_{AE}(t,z) =  i \sum_\lambda \int_0^\infty d \omega \ \sqrt{\frac{\omega_{\lambda}}{\omega}}g^*_{\lambda} \big(\chi^*_{\lambda}(\omega)\hat{\Phi}^\dag_{\lambda \omega -}(t,z)
    +  \chi_{\lambda}(\omega)\hat{\Phi}_{\lambda \omega +}(t,z) \big).
\end{align}

Assuming that the scattering rate is much larger than the characteristic rate of change of the phonon envelope, the decay of the plasmons is so rapid that the acoustoelectric interactions are effectively local in space. In this limit, $B(t-\tau, z- v_d\tau)$ under the integral in Eqs. \eqref{Eq: HL-phonon-2}  can be approximated by $B(t-\tau, z- v_d\tau)\approx B(t-\tau, z)$. Since the dominant contribution to the integral come from $\tau < 2/\gamma$, this approximation is equivalent to neglecting a shift in position $\Delta z$ of order $v_d/\gamma$. For a typical system, $v_d \sim 5000$ m/s and $\gamma \sim (2\pi) 1$ THz, corresponding with a neglected spatial shift of $\Delta z < 1$ nm, which is much smaller than typical phonon wavelengths for systems modeled here. 

Using this approximation and including intrinsic phonon losses ($\Gamma$) and noise ($\xi$) (i.e., fluctuations and dissipation that drive the uncoupled phonons into thermal equilibrium), the Heisenberg-Langevin equations in the Fourier domain reduce to 
\begin{eqnarray}
\label{Eq: Heisenberg-Langevin}
    -i\omega{B}(\omega,z) =  -\bigg(i\big(\Omega+\Delta\Omega_{AE}(\omega)\big) +\frac{1}{2}\big(\Gamma-v_g G_{AE}(\omega)\big)\bigg)B(\omega,z) - v_g \partial_z B(\omega,z) + \underbrace{\xi(\omega,z) + \xi_{AE}(\omega,z)}_{\xi_{eff}(\omega,z)}.
\end{eqnarray}
\end{widetext}
Here, the backreaction of plasmons yields a shift in the phonon resonant frequency $\Delta\Omega_{AE}$ and gain per unit length $G_{AE}$ defined by 
\begin{align}
    & \Delta\Omega_{AE}(\omega) =  \sum_\lambda  \frac{2|{g}_\lambda|^2 {\omega}_\lambda((v_d q_m-\omega)^2-\omega_\lambda^2)}{((v_d q_m-\omega)^2-\omega_\lambda^2)^2+\gamma^2(v_d q_m-\omega)^2}
    \\
& G_{AE}(\omega) = \frac{1}{v_g} \sum_\lambda  \frac{4|{g}_\lambda|^2 {\omega}_\lambda \gamma(v_d q_m-\omega)}{((v_d q_m-\omega)^2-\omega_\lambda^2)^2+\gamma^2(v_d q_m-\omega)^2}. 
\end{align}

Key features of acoustoelectric systems are contained in the equation for $G_{AE}$. For $\omega \approx \Omega$, $G_{AE}$ leads to phonon gain when $v_d q_m > \Omega$, or equivalently when the drift velocity exceeds the mechanical phase velocity. As expected, $G_{AE} < 0$ for $v_d q_m < \Omega$, leading to excess loss when the speed of sound is greater than $v_d$. 

Equation \eqref{Eq: Heisenberg-Langevin} also shows how mechanical fluctuations are modified, the mechanical Langevin force  that drives the system into thermal equilibrium $\xi$ is augmented by additive noise, i.e., $\xi \to \xi_{eff}=  \xi + \xi_{AE}$. Assuming $\Omega \gg \Gamma$, the Langevin force $\xi$ is approximately given by 
\begin{align}
    \xi(t,z) =  \int_0^\infty d \omega \ \xi_{\omega}(z) e^{-i\omega t}
\end{align}
where $\langle \xi^\dag_{\omega}(z) \xi_{\omega'}(z') \rangle =  (\Gamma/2\pi) n(\omega) \delta(\omega-\omega')\delta(z-z')$ and $n(\omega)$ is the Bose-Einstein thermal occupation number. With these results, the two-point correlation function for $\xi$ is given by
\begin{align}
\label{Eq: xi xi 1}
    &\!\! \langle \xi^\dag(t,z) \xi(t',z') \rangle \! = \! \int_0^\infty  \frac{d\omega}{2\pi}  \Gamma n(\omega) e^{i\omega(t-t')}\delta(z\!-\!z') \\
    \label{Eq: xi xi 2}
    & \!\! \langle \xi(t,z) \xi^\dag(t',z') \rangle \!=\! \int_0^\infty \frac{d\omega}{2\pi}  \Gamma \big(n(\omega)\!+\!1\big)e^{-i\omega(t-t')}\delta(z\!-\!z'),
\end{align}
which (1) generalizes beyond a simple white noise model, (2) more accurately captures the frequency dependence of the thermal occupation, and (3) yields results that are consistent with the fluctuation-dissipation theorem in the limit $\Omega \gg \Gamma$ \cite{callen1951irreversibility}. 

\begin{figure*}[t]
    \centering
    \includegraphics[width=0.95\linewidth]{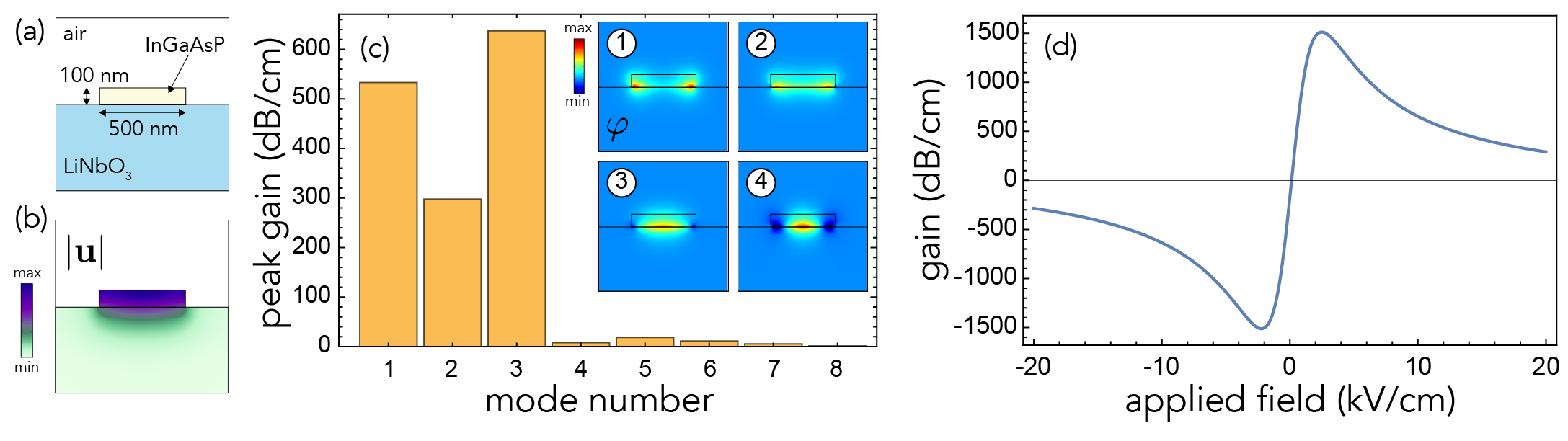}
    \caption{Simulation of acoustoelectric gain in an InGaAsP on LiNbO$_3$ waveguide structure based on the analytical results presented here. (a) Waveguide cross-sectional geometry. (b) Displacement magnitude of the $(2\pi)8.519$ GHz guided mechanical mode considered that is utilized for this acoustoelectric gain analysis. (c) Peak acoustoelectric gain for the various plasmon modes. Inset: electric potential for the first four plasmon modes. (d) acoustoelectric gain for phonon mode shown in (b) as a function of applied electric field. The carrier drift velocity is given by $v_d = \mu E$ where the mobility is $\mu = 2000$ cm$^2$/kV. The materials properties of LiNbO$_3$ and InGaAsP used in  this analysis are summarized in Appendix \ref{App: Materials Properties}.}
    \label{Fig: GAE}
\end{figure*}

To illustrate the utility of these results, we calculate the gain for an acoustoelectric waveguide structure comprised of LiNbO$_3$ half-space and a rib of InGaAsP shown in Fig. \ref{Fig: GAE}(a). The rib supports a guided Rayleigh-like surface wave (Fig. \ref{Fig: GAE} (b)) which can couple to a family of surface plasmon modes (see inset of \ref{Fig: GAE} (c)) in a manner relating to the acoustoelectric overlap described by Eq. \eqref{Eq: AE coupling}. The gain (loss) produced by acoustoelectric coupling is shown as a function of applied field (i.e., proportional to the free-carrier drift velocity).  

\section{Phonon noise}
\label{Sec: Phonon Noise}
In this section, we calculate the power spectrum for phonon noise that includes contributions from phonon losses and fluctuations that accompany acoustoelectric coupling (see Ref. \cite{kino1973noise,hackett2023non} for noise analysis including charge trapping and diffusion). In the Fourier domain, the formal solution to the phonon envelope equation is 
\begin{align}
\label{B-for-sol}
    B(\omega,z) = \hat{B}(\omega,0)e^{-\Lambda z}+ \frac{1}{v_g} \int_0^z dz_1 e^{-\Lambda(z-z_1)} \xi_{eff}(\omega,z_1). 
\end{align}
where $\Lambda \equiv \big(i(\Omega+\Delta\Omega_{AE} -\omega)+(\Gamma-v_g G_{AE})/2\big)/v_g$, and $\hat{B}(\omega,0)$ represents the phonon noise entering the region $z > 0$.
The power spectrum for the phonon noise can be obtained from the ensemble average

\begin{widetext}
\begin{align}
\label{Eq: B-power-spectrum}
   S_B(\omega,z) = & \frac{\langle B^\dag(\omega,z)  B(\omega',z) \rangle}{2\pi \delta(\omega-\omega')} 
   \\ = &
   \frac{\langle \hat{B}^\dag(\omega,0)  \hat{B}(\omega',0) \rangle}{2\pi \delta(\omega-\omega')}e^{-2 {\rm Re}[\Lambda] z}
   +
   \frac{1}{v_g^2}\int_0^z dz_1 
   \int_0^z dz_2 \
   e^{-\Lambda^*(z-z_1)-\Lambda(z-z_2)}
    \frac{\langle \xi_{eff}^\dag(\omega,z_1)  \xi_{eff}(\omega',z_2) \rangle}{2\pi \delta(\omega-\omega')}   
   .\nonumber
\end{align}
\end{widetext}

To obtain the noise produced by acoustoelectric interactions, we need to calculate the two-point correlation function of $\hat{\Phi}_{\lambda \omega \pm}$. Recall that this operator is the homogeneous solution to the plasmon operator which is given by
\begin{eqnarray}
\label{Eq: Phi-Lan-T}
    \hat{\Phi}_{\lambda \omega \pm}(t,z) = \int \frac{dq}{\sqrt{2\pi}}
    e^{-i[(\omega + v_d q)t +(\pm q_m-q)z]} \hat{a}_{\lambda q \omega}.
\end{eqnarray}
Here, $\hat{a}_{\lambda q \omega}$($\hat{a}_{\lambda q \omega}$) are the Schrodinger picture annihilation (creation) operators for the plasmon-reservoir system.

In the presence of a drift current, the plasmonic modes are in a nonequilibrium steady state, and therefore there are no general theorems that can be used to calculate the plasmon fluctuations. We will estimate these fluctuations by assuming the form of the occupation number $n_{\lambda q \omega}$ (i.e., the number of plasmons in a given mode) that derives from the expectation value $\langle \hat{a}^\dag_{\lambda q \omega} \hat{a}_{\lambda' q' \omega'} \rangle$. Assuming that thermal equilibrium is established before the drift current is initiated, the thermal occupation is given by $n(\omega) = (\exp\{\hbar \omega \beta\}-1)^{-1}$ with $\beta = (k_B T)^{-1}$
and the expectation value $\langle \hat{a}^\dag_{\lambda q \omega} \hat{a}_{\lambda' q' \omega'} \rangle$ is given by
\begin{align}
    & \langle {\hat{a}^{\dag}}_{\lambda q \omega} \hat{a}_{\lambda' q' \omega'} \rangle = n(\omega) \delta_{\lambda \lambda'}\delta(q-q') \delta(\omega-\omega' )\\
   &  \langle {\hat{a}}_{\lambda q \omega} \hat{a}^{\dag}_{\lambda' q' \omega'} \rangle = \big(n(\omega) +1\big)\delta_{\lambda \lambda'}\delta(q-q') \delta(\omega-\omega').
\end{align}
Under these assumptions, the two-point correlation functions for $\xi_{AE}$ is given by 
\begin{align}
\label{Eq: xi_AE-corr}
    \frac{\langle \xi_{AE}^\dag(\omega,z_1)  \xi_{AE}(\omega',z_2) \rangle}{2\pi \delta(\omega-\omega')}& \!=\!  \sum_{\lambda}\!\int dq  \frac{ \omega_\lambda |g_\lambda|^2}{(v_d q\!-\! \omega)} e^{i(q_m\!-\!q)(z_1\!-\!z_2)}
    \nonumber \\
     &\times |\chi_\lambda(v_d q\!-\! \omega)|^2
    \big( n(v_d q\!-\! \omega)\!+\!1 \big)  
\end{align}
where the identity $-n(-\omega) = n(\omega) + 1$ has been used.

Assuming that no acoustoelectric interactions occur in the region $z < 0$, the injected noise is given by $\langle \hat{B}^\dag(\omega,0)  \hat{B}(\omega',0) \rangle = (n(\omega)/v_g) 2\pi \delta(\omega-\omega')$ (see Appendix \ref{App: injected phonon noise}). Inserting this result and Eqs. \eqref{Eq: xi xi 1} \& \eqref{Eq: xi_AE-corr} into Eq. \eqref{Eq: B-power-spectrum} and evaluating the $z_1$ and $z_2$ integrals, the phonon power spectrum is given by 

\begin{widetext}
    \begin{align}
     S_{B}(\omega,z) = 
    \frac{1}{v_g}  & \frac{\alpha\!-\!G_{AE}(\omega)e^{-(\alpha-G_{AE}(\omega))z}}{\alpha-G_{AE}(\omega)}   n(\omega)  
    \!+\!
   \frac{1}{v_g^2}\sum_{\lambda}\!\int \! dq \bigg| \frac{e^{-(\Lambda-i(q_m-q))z}\!-\!1}{\Lambda-i(q_m-q)} \bigg|^2
    \frac{ \omega_\lambda |g_\lambda|^2}{(v_d q\!-\! \omega)}  |\chi_\lambda(v_d q\!-\! \omega)|^2\big( n(v_d q\!-\! \omega)\!+\!1 \big) 
 \end{align} 
 where $\alpha = \Gamma/v_g$. Under the condition that $\gamma/v_d \gg {\rm max}(2\pi/z,\alpha -G_{AE})$, the term $|(e^{-(\Lambda-i(q_m-q))z}-1)/(\Lambda-i(q_m-q))|^2$ in the integrand above is sharply peaked and can be approximated as 
 \begin{align}
     \bigg| \frac{e^{-(\Lambda-i(q_m-q))z}-1}{\Lambda-i(q_m-q)} \bigg|^2 \approx \frac{2\pi}{\alpha -G_{AE}(\omega)}\bigg[1-e^{-(\alpha -G_{AE}(\omega))z} \bigg]\delta(q-q_m+(\Omega+\Delta\Omega_{AE} -\omega)/v_g).
 \end{align}
  \end{widetext}
Taking $\omega = \Omega + \Delta\Omega_{AE} \approx \Omega$ enables a drastic simplification of the power spectrum given approximately by  
\begin{align}
\label{Eq: phonon power spectrum}
    S_B(\Omega,z)  
    =  \frac{1}{v_g}n(\Omega) - & \frac{1}{v_g}\frac{1-e^{-(\alpha-G_{AE}(\Omega))z}}{\alpha-G_{AE}(\Omega)} 
    \\ & \times G_{AE}(\Omega) \big( n(\Omega-v_d q_m)- n(\Omega) \big) \nonumber 
\end{align}

Two limits show the physical consistency of this result. In the absence of acoustoelectric coupling, i.e, $G_{AE} \to 0$, or when the drift current is set to $0$, the phonon power spectrum reduces to the expected thermal equilibrium result $S_B(\Omega,z) \to n(\Omega)/v_g$.  

Using this power spectrum, we can compute the noise factor $F$ (or the noise figure $NF = 10 \log F)$  for the amplifying region. As a key figure of merit for amplifier performance, the noise factor quantifies the degradation of the signal-to-noise ratio upon amplification and through additive noise. $F$ is defined as the ratio of the injected signal-to-noise $SNR_{i}$ to the signal to noise after amplification $SNR_o$ given by
\begin{eqnarray}
    F =  \frac{{\rm SNR}_i}{{\rm SNR}_o}
\end{eqnarray}
with ${\rm SNR}_i = S_i/(n(\Omega) \Delta\omega /v_g)$ where $S_i$ is the injected signal power and $\Delta \omega$ is the detection bandwidth, and ${\rm SNR}_o = S_o/(S_B(\Omega,z )\Delta\omega)$ where $S_o = S_i \exp\{-(\alpha-G_{AE}(\Omega))z \}$ is the signal power exiting the amplifier. For the acoustoelectric system, the noise factor is 
\begin{align}
    F = 1+\frac{e^{(\alpha-G_{AE}(\Omega))z}-1}{\alpha-G_{AE}(\Omega)}  \bigg(\alpha-G_{AE}(\Omega)\frac{n(\Omega-v_d q_m)}{n(\Omega)} \bigg).
\end{align}

\begin{figure}
    \centering
    \includegraphics[width=0.95\linewidth]{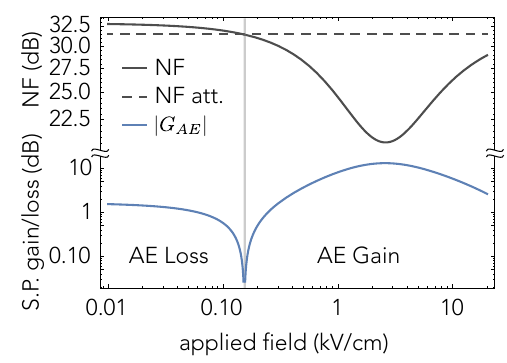}
    \caption{Noise figure (solid gray) and single pass (S.P.) acoustoelectric gain/loss (blue) vs. applied field. The dashed line is the noise figure for the system in the absence of acoustoelectric coupling. The system parameters are take from Fig. \ref{Fig: GAE} and Tab. \ref{tab: parameters}. The device length $z$ is set at $100 \mu$m, the loss per unit length $\alpha$ is set to $3116$ dB/cm, and selected temperature is 300 K.}
    \label{fig: noise figure}
\end{figure}

Figure \ref{fig: noise figure} shows the noise figure versus applied electric field for a $100 \mu$m long waveguide of cross-section shown in Fig. \ref{Fig: GAE}(a). As the acoustoelectric gain increases, the concomitant amplification leads to a decrease in the noise factor, below the level set by the mechanical attenuation $\alpha$ in the amplifying region.  

\section{Conclusion}
In this work, we have developed a quantum field theoretic framework for acoustoelectric interactions in waveguide-like systems of arbitrary cross-section. By formulating the dynamics within an open quantum systems approach, we derived a description of plasmon–phonon coupling that incorporates dissipation, noise, and the role of drift currents. Our treatment captures both bulk and surface plasmon modes, with special emphasis on surface modes that dominate in semiconductor-piezoelectric heterostructures.

The resulting Heisenberg-Langevin equations reveal how drift currents Doppler-shift plasmonic resonances and modify the phonon noise spectrum, providing new insights into the nonequilibrium fluctuations that arise in these systems. Importantly, our model recovers known limits while also extending them to account for complex geometries, strong dissipation, and nontrivial noise dynamics. The derived expressions for gain, frequency shifts, and noise power spectra provide quantitative tools for evaluating performance metrics such as the noise factor, thereby enabling rigorous analysis of acoustoelectric amplifiers and oscillators. Looking to the future, this work provides a foundation for optimizing existing architectures and exploring new operational regimes where acoustoelectric interactions can be harnessed for low-noise amplification, nonreciprocal transport, and hybrid quantum technologies.
\section*{Funding Information}
This material is based upon work supported by the Laboratory Directed Research and Development program at
Sandia National Laboratories. Sandia National Laboratories is a multi-program laboratory managed and operated by National Technology and Engineering Solutions
of Sandia, LLC., a wholly owned subsidiary of Honeywell
International, Inc., for the U.S. Department of Energy’s
National Nuclear Security Administration under contract
DE-NA-0003525. This paper describes objective technical results and analysis. Any subjective views or opinions
that might be expressed in the paper do not necessarily
represent the views of the U.S. Department of Energy or
the United States Government. 

\section{Acknowledgements}
ROB dedicates this work to Rosemary Bean who has been a continuous source of inspiration. 


\appendix

\section{Variations of surface integral Lagrangians}
\label{App: variation of surface Lagrangian}
Here, we show how variations of $L_S$ are taken to arrive at correct equations of motion. It is the kinetic energy term $T_S$ given by 
\begin{align}
T_S = \oint_{\partial V_{in}} da \ \frac{1}{2}m n_0 \dot{\psi} \frac{\partial \dot{\psi}}{\partial x_\perp}
\end{align}
that is most critical. Although the velocity potential for the surface modes satisfies the Laplace equation $-\nabla^2 \dot{\psi} = 0$, variations of $\dot{\psi}$ may not. This can be correctly accounted for by expressing $T_S$ as a volume integral 
\begin{align}
T_S = \int_{ V_{in}} d^3 x \ \frac{1}{2}m n_0 (\nabla \dot{\psi})^2 
\end{align}
taking the variation, and then restricting to the case where $-\nabla^2 \dot{\psi} = 0$. Therefore, variation of $T_S$
\begin{align}
    \delta T_S = & -\int_{ V_{in}} d^3 x \ m n_0  \nabla \dot{\psi} \cdot \nabla \delta \dot{\psi}
    \nonumber
    \\ 
    = & -\oint_{\partial V_{in}} da \ m n_0 \delta \dot{\psi} \frac{\partial \dot{\psi}}{\partial x_\perp} + \int_{ V_{in}} d^3 x \ m n_0   \delta \dot{\psi} \cancelto{0}{\nabla^2  \dot{\psi}}.
\end{align}
This shows that 
\begin{align}
  \left.  \frac{\delta T_S}{\delta \dot{\psi}({\bf x}) }\right|_{{\bf x}\in \partial V_{in}} = -m n_0 \frac{\partial \dot{\psi}}{\partial x_\perp}.
\end{align}

\section{Orthogonality and completeness of the surface plasmon modes}
\label{App: orthogonality}
Here, we show that $\oint_{\partial A_{in}} d\ell \  
            \phi^*_{\lambda q}  
          \frac{\partial \phi^{in}_{\lambda q}}{\partial x_\perp} \geq 0
          $, a key ingredient to derive the orthogonality relations for the surface plasmon modes. 
Using Eq. \eqref{SL2}, we can show
\begin{align}
 \label{orth_rel_2}
     \oint_{\partial A_{in}} d\ell \  
            \phi^*_{\lambda q}  
          \frac{\partial \phi^{in}_{\lambda q}}{\partial x_\perp}
        & =  \frac{1}{2} \oint_{\partial A_{in}} d\ell \  
          \frac{\partial |\phi^{in}_{\lambda q}|^2}{\partial x_\perp}
      \\
        & =   \frac{1}{2}\int_{A_{in}} da \ \nabla_\perp^2 |\phi^{in}_{\lambda q}|^2
        \\
        & =   \int_{A_{in}} da \ (q^2 |\phi^{in}_{\lambda q}|^2 + |\nabla_\perp \phi^{in}_{\lambda q}|^2) \geq 0
 \end{align}
 where the divergence theorem is used in the second step and Eq. \eqref{eig_func} in the last step. For inner products of the form $\oint_{\partial A_{in}} d\ell \  
            \phi_{\lambda' q}  
          \frac{\partial \phi^{in}_{\lambda q}}{\partial x_\perp} 
          $,
          Eq. \eqref{orth_rel} becomes
 \begin{eqnarray}
 \label{orth_rel_3}
       (\omega_{\lambda'q}^{2}-\omega_{\lambda q}^{2}) \oint_{\partial A_{in}} d\ell \ 
            \phi_{\lambda' q}  
          \frac{\partial \phi^{in}_{\lambda q}}{\partial x_\perp}
        = 0
 \end{eqnarray}
 which can be satisfied by demanding $\oint_{\partial A_{in}} d\ell \  
            \phi_{\lambda' q}  
          \frac{\partial \phi^{in}_{\lambda q}}{\partial x_\perp} 
          = 0$. Utilizing the orthonormality relation Eq. \eqref{orth_rel}, one can show
\begin{eqnarray}
    \delta^2({\bf x}-{\bf x}') = \sum_\lambda \int \frac{dq}{2\pi} \phi^*_{\lambda q} ({\bf x}) \frac{\partial \phi^{in}_{\lambda q} ({\bf x}')}{\partial x_\perp} e^{i q(z-z')}
\end{eqnarray}
for points ${\bf x}, {\bf x}' \in \partial A_{in}$.

\section{Diagonalization of open systems Hamiltonian}
\label{App: diagonalization}
In this section we show how the decomposition of the system and bath operators given by Eqs. \eqref{Eq:Sys-x}-\eqref{Eq: bath-p} diagonalizes the open system Hamiltonian $H$. To begin, consider the decomposition of  each term in the Hamiltonian $H_{\lambda q}$ 
\begin{widetext}
\begin{align}
& \frac{p^\dag_{\lambda q} p_{\lambda q}}{2m} = \int_0^\infty \!\!\!\! d\omega \int_0^\infty \!\!\!\! d\omega' \frac{\hbar\sqrt{\omega \omega'}}{4}[\chi_{\lambda q}^*(\omega)a^\dag_{\lambda q \omega}-\chi_{\lambda q}(\omega)a_{\lambda,-q \omega}][\chi_{\lambda q}(\omega')a_{\lambda q \omega'}-\chi_{\lambda q}^*(\omega')a^\dag_{\lambda,-q \omega'}]
\\
& i v_d q \ p_{\lambda q} x_{\lambda q}^\dag = \int_0^\infty \!\!\!\! d\omega \int_0^\infty \!\!\!\! d\omega' \frac{\hbar v_d q}{2}[\chi_{\lambda q}(\omega)a_{\lambda q \omega}-\chi_{\lambda q}^*(\omega)a^\dag_{\lambda,-q \omega}][\chi_{\lambda q}^*(\omega')a^\dag_{\lambda q \omega'}+\chi_{\lambda q}(\omega')a_{\lambda,-q \omega'}]
\\
& \frac{m \omega_{\lambda q}^2}{2}   x^\dag_{\lambda q} x_{\lambda q} = \int_0^\infty \!\!\!\! d\omega \int_0^\infty \!\!\!\! d\omega' \frac{\hbar \omega_{\lambda q}^2}{4 \sqrt{\omega \omega'}} [\chi_{\lambda q}^*(\omega)a^\dag_{\lambda q \omega}+\chi_{\lambda q}(\omega)a_{\lambda,-q \omega}][\chi_{\lambda q}(\omega')a_{\lambda q \omega'}+\chi_{\lambda q}^*(\omega')a^\dag_{\lambda,-q \omega'}]
\\
& \int_0^\infty \!\!\!\!  d\nu \frac{p^\dag_{\lambda q \nu} p_{\lambda q \nu}}{2m} = \int_0^\infty \!\!\!\! d\omega \int_0^\infty \!\!\!\! d\omega' \int_0^\infty \!\!\!\! d\nu \frac{\hbar\sqrt{\omega \omega'}}{4}[\chi_{\lambda q\nu}^*(\omega)a^\dag_{\lambda q \omega}-\chi_{\lambda q\nu}(\omega)a_{\lambda,-q \omega}][\chi_{\lambda q\nu}(\omega')a_{\lambda q \omega'}-\chi_{\lambda q \nu}^*(\omega')a^\dag_{\lambda,-q \omega'}]
\\
& \int_0^\infty \!\!\!\!  d\nu \ i v_d q  p_{\lambda q \nu} x_{\lambda q \nu}^\dag = \int_0^\infty \!\!\!\! d\omega \int_0^\infty \!\!\!\! d\omega' \int_0^\infty \!\!\!\! d\nu  \frac{\hbar v_d q}{2}[\chi_{\lambda q\nu}(\omega)a_{\lambda q \omega}-\chi_{\lambda q\nu}^*(\omega)a^\dag_{\lambda,-q \omega}][\chi_{\lambda q\nu}^*(\omega')a^\dag_{\lambda q \omega'}+\chi_{\lambda q\nu}(\omega')a_{\lambda,-q \omega'}]
\\
& \int_0^\infty \!\!\!\! d\nu \frac{m \nu^2}{2}  x^\dag_{\lambda q \nu} x_{\lambda q \nu} = \! \int_0^\infty \!\!\!\! d\omega \int_0^\infty \!\!\!\! d\omega' \!\! \int_0^\infty \!\!\!\! d\nu \frac{\hbar \nu^2}{4 \sqrt{\omega \omega'}} [\chi_{\lambda q\nu}^*(\omega)a^\dag_{\lambda q \omega}+\chi_{\lambda q\nu}(\omega)a_{\lambda,-q \omega}][\chi_{\lambda q\nu}(\omega')a_{\lambda q \omega'}+\chi_{\lambda q\nu}^*(\omega')a^\dag_{\lambda,-q \omega'}]
\\
& \int_0^\infty \!\!\!\! d\nu \frac{m \nu^2 c_\nu}{2}  x^\dag_{\lambda q \nu} x_{\lambda q } = \! \int_0^\infty \!\!\!\! d\omega \int_0^\infty \!\!\!\! d\omega' \!\! \int_0^\infty \!\!\!\! d\nu \frac{\hbar G_\nu}{4 \sqrt{\omega \omega'}} [\chi_{\lambda q\nu}^*(\omega)a^\dag_{\lambda q \omega}+\chi_{\lambda q\nu}(\omega)a_{\lambda,-q \omega}][\chi_{\lambda q}(\omega')a_{\lambda q \omega'}+\chi_{\lambda q}^*(\omega')a^\dag_{\lambda,-q \omega'}] 
\\
& \int_0^\infty \!\!\!\! d\nu \frac{m \nu^2 c_\nu^2}{2}  x^\dag_{\lambda q} x_{\lambda q } = \! \int_0^\infty \!\!\!\! d\omega \int_0^\infty \!\!\!\! d\omega' \!\! \int_0^\infty \!\!\!\! d\nu \frac{\hbar }{4 \sqrt{\omega \omega'}} \frac{G_\nu^2}{\nu^2} [\chi_{\lambda q}^*(\omega)a^\dag_{\lambda q \omega}+\chi_{\lambda q}(\omega)a_{\lambda,-q \omega}][\chi_{\lambda q}(\omega')a_{\lambda q \omega'}+\chi_{\lambda q}^*(\omega')a^\dag_{\lambda,-q \omega'}]
\end{align}
To see that the Hamiltonian diagonalizes with the new operators $a_{\lambda q \omega}$, consider $H_1$, the portion of the Hamiltonian containing $a_{\lambda q \omega}^\dag a_{\lambda q \omega'}$. Collecting the relevant terms from the expansions above, we find
\begin{eqnarray}
H_1 = \sum_\lambda \int dq \int_0^\infty d\omega \int_0^\infty d\omega'
\frac{\hbar}{4\sqrt{\omega \omega'}} \bigg[ &&
(\omega \omega' + \omega_{\lambda q}^2 + 2\sqrt{\omega \omega'}v_d q) \chi^*_{\lambda q}(\omega)\chi_{\lambda q}(\omega')
\nonumber \\
&&
+
\int_0^\infty d\nu (\omega \omega' + \nu^2 + 2\sqrt{\omega \omega'}v_d q) \chi^*_{\lambda q \nu}(\omega)\chi_{\lambda q \nu}(\omega')
\nonumber \\
&&
-2\int_0^\infty \! d \nu G_\nu \chi^*_{\lambda q \nu}(\omega)\chi_{\lambda q }(\omega')
+\!
\int_0^\infty \! d\nu \frac{G_\nu^2}{\nu^2}\chi^*_{\lambda q }(\omega)\chi_{\lambda q}(\omega')
\bigg]a^\dag_{\lambda q \omega}a_{\lambda q \omega'} \quad
\end{eqnarray}
Using the definitions of the susceptibilities (Eqs. \eqref{Eq: susc-1}-\eqref{Eq: susc-2}), the integrals over $\nu$ within the square brackets above can be simplified
\begin{eqnarray}
\label{Eq: int-rel-1}
   && \int_0^\infty d\nu \ \nu^2 \chi^*_{\lambda q \nu}(\omega)\chi_{\lambda q \nu}(\omega') = G_\omega \chi_{\lambda q}(\omega') +(-i \gamma \omega +\Delta-{\omega'}^2)\chi^*_{\lambda q}(\omega) \chi_{\lambda q}(\omega')+{\omega'}^2\delta(\omega-\omega')
   \\
   \label{Eq: int-rel-2}
  && -2\int_0^\infty d \nu \ G_\nu \chi^*_{\lambda q \nu}(\omega)\chi_{\lambda q }(\omega') = -2 G_\omega \chi_{\lambda q}(\omega') - 2(-i \gamma \omega +\Delta)\chi^*_{\lambda q}(\omega) \chi_{\lambda q}(\omega')
\end{eqnarray}
where $\Delta = \int_0^\infty d\nu \ G_\nu^2/\nu^2 $.
Using Eqs. \eqref{Eq: int-rel-1}-\eqref{Eq: int-rel-2}, the Lippmann-Schwinger orthogonality conditions (Eqs. \eqref{Eq: LS-1}-\eqref{Eq: LS-4}) and the definitions of the susceptibilities, $H_1$ reduces to
\begin{eqnarray}
    H_1 = \sum_\lambda \int dq \int_0^\infty d\omega \int_0^\infty d\omega' \bigg[ \frac{\hbar}{2}(\omega +v_d q)\delta(\omega - \omega')
+\frac{\hbar}{4\sqrt{\omega \omega'}}(\omega^2-{\omega'}^2)\chi^*_{\lambda q}(\omega)\chi_{\lambda q}(\omega')\bigg]
    a^\dag_{\lambda q \omega}a_{\lambda q \omega'}.
\end{eqnarray}
In analogy with the simplification of $H_1$ above, the remaining terms in the open systems Hamiltonian can be derived. The component of the Hamiltonian $H_2$ containing $a_{\lambda q \omega}a^\dag_{\lambda q \omega}$ reduces to 
\begin{eqnarray}
    H_2 = \sum_\lambda \int dq \int_0^\infty d\omega \int_0^\infty d\omega' \bigg[ \frac{\hbar}{2}(\omega +v_d q)\delta(\omega - \omega')
-\frac{\hbar}{4\sqrt{\omega \omega'}}(\omega^2-{\omega'}^2)\chi^*_{\lambda q}(\omega)\chi_{\lambda q}(\omega')\bigg]
    a_{\lambda q \omega'}a^\dag_{\lambda q \omega}.
\end{eqnarray}
A tedious calculation shows that all remaining terms vanish (i.e., those containing $a_{\lambda q \omega}a_{\lambda q \omega}$ and $a^\dag_{\lambda q \omega}a^\dag_{\lambda q \omega}$). Adding $H_1$ and $H_2$, and using the commutation relation Eqs. \eqref{Eq:CR-1} the total Hamiltonian is given by 
\begin{eqnarray}
    H = \frac{1}{2}\sum_\lambda \int dq \int_0^\infty d\omega \ \hbar(\omega +v_d q)(a^\dag_{\lambda q \omega}a_{\lambda q \omega}+
    a_{\lambda q \omega}a^\dag_{\lambda q \omega})
\end{eqnarray}
\end{widetext}

\section{Simplification of the Heisenberg-Langevin equations}
\label{Sec: Simplification HE} 
Here, we explicitly show hidden steps in the derivation of the Heisenberg-Langevin equations. Inserting Eqs. \eqref{Eq: plasmon_right_sol} \& \eqref{Eq: plasmon_left_sol} into Eq. \eqref{Eq: HOM-phonon} leads to the following effective phonon equation of motion
\begin{widetext}
\begin{align}
\label{Eq: HL-phonon}
 \dot{B}(t,z) = 
    -i\Omega B(t,z) - v_g \partial_z B(t,z) 
    + \sum_\lambda \int_0^\infty d\tau & \int_0^\infty d\omega \frac{\omega_\lambda}{\omega}  |g_\lambda|^2|\chi_\lambda(\omega)|^2 e^{-iv_dq_m\tau} \bigg(e^{i\omega\tau}  
    -e^{-i\omega\tau}\bigg)B(t-\tau,z-v_d \tau) \nonumber \\ & + i \underbrace{\sum_\lambda \int_0^\infty d \omega \ \sqrt{\frac{\omega_{\lambda}}{\omega}}g^*_{\lambda} \big(\chi^*_{\lambda}(\omega)\hat{\Phi}^\dag_{\lambda \omega -}(t,z)
    +  \chi_{\lambda}(\omega)\hat{\Phi}_{\lambda \omega +}(t,z) \big)}_{\xi_{AE}(t,z)}. 
\end{align}
By performing the $\omega$-integral
\begin{align}
  \int_0^\infty d\omega \ \frac{1}{\omega} & |\chi_\lambda(\omega)|^2  \bigg(e^{i\omega\tau}-  e^{-i\omega\tau}\bigg)  = \frac{1}{\tilde{\omega}_\lambda}e^{-\frac{1}{2}\gamma \tau}\bigg(
 e^{i \tilde{\omega}_\lambda \tau} - e^{-i \tilde{\omega}_\lambda \tau} \bigg)
\end{align}
the Heisenberg-Langevin equations reduce to Eq. \eqref{Eq: HL-phonon-2}
\end{widetext}

\section{Materials properties of LiNbO$_3$ and InGaAsP used in acoustoelectric gain simulations}
\label{App: Materials Properties}

The materials properties for lithium niobate and InGaAsP used for the simulations depicted in Fig. \ref{Fig: GAE} are given in the following tables
\begin{table}[h]
    \centering
    \begin{tabular}{c|c|c}
       Y & Young's Modulus  & 170 \ {\rm GPa}  \\
       $\nu$ & Poisson ratio    & 0.25 \\
       $\rho$ & density          & 4700  kg/m$^3$ \\    
       $\varepsilon$ & permittivity & 29.2 \\       
       \end{tabular}
    \caption{Materials properties of lithium niobate used in acoustoelectric gain simulations.}
    \label{tab: parameters}
\end{table}

\begin{table}[h]
    \centering
    \begin{tabular}{c|c|c}
       $c_p$ & pressure wave speed  & 4371 \ m/s  \\
       $c_s$ & shear wave speed     & 2900 \ m/s \\
       $\rho$ & density          & 5232 \ kg/m$^3$ \\    
       $\omega_p$ & plasma frequency & $(2\pi)1.04$ \ THz \\
       $\varepsilon_{in}$ & permittivity & 13.3 \\
       $\gamma$ & scattering rate & $(2\pi)$2.516 \  THz \\
       $\mu$ & electron mobility & 2000 ${\rm cm}^2/{\rm kV}$
       \end{tabular}
    \caption{Materials properties of InGaAsP used in acoustoelectric gain simulations.}
    \label{tab:placeholder}
\end{table}
\section{Injected phonon noise}
\label{App: injected phonon noise}
This appendix calculates the noise injected into the acoustoelectric region that is modeled by $\hat{B}(\omega,0)$. When acoustoelectric interactions are neglected ($G_{AE} \to 0$), the solution for the phonon envelope at $z=0$ is given by 
\begin{eqnarray}
\label{B-injected}
    \hat{B}(\omega,0) \! = \! \frac{1}{v_g} \! \int_{-\infty}^0 \!\! dz_1 e^{-\frac{1}{v_g}(i(\Omega-\omega) +\Gamma/2)(z-z_1)} \xi(\omega,z_1),  \ \ 
\end{eqnarray}
where the system is assumed to be semi-infinite. Using the correlation properties of $\xi$ given by Eq. \eqref{Eq: xi xi 1}, the power spectrum of $\hat{B}(\omega,0)$ is given by
\begin{widetext}
\begin{align}
  \frac{\langle \hat{B}^\dag(\omega,0)  \hat{B}(\omega',0) \rangle}{2\pi \delta(\omega-\omega')} & =  \frac{1}{v_g^2} \int_{-\infty}^0 dz_1 \int_{-\infty}^0 dz_2 \ e^{-\frac{1}{v_g}(-i(\Omega-\omega') +\Gamma/2)(z-z_1)} 
  e^{-\frac{1}{v_g}(i(\Omega-\omega) +\Gamma/2)(z-z_2)}  \frac{\langle \xi^\dag(\omega,z_1)  \xi(\omega',z_2) \rangle}{2\pi \delta(\omega-\omega')}
 \nonumber  \\
  & = \frac{1}{v_g^2} \int_{-\infty}^0 dz_1 \int_{-\infty}^0 dz_2 \ e^{\frac{1}{v_g}\Gamma(z-z_1)} 
   \Gamma n(\omega) \delta(z_1-z_2) \nonumber 
  \\
  & =
  \frac{1}{v_g} n(\omega).
\end{align}
\end{widetext}
This result models the injection of thermal noise into the amplifying region.


\bibliography{cites}



\end{document}